\documentclass[conference, letterpaper]{IEEEtran}
\IEEEoverridecommandlockouts
\usepackage{booktabs}
\usepackage{balance}
\usepackage{tabularx}
\usepackage{cite}
\usepackage{amsmath,amssymb,amsfonts}
\usepackage{algorithmic}
\usepackage{graphicx}
\usepackage{tabularx}
\usepackage{textcomp}
\usepackage[framemethod=TikZ]{mdframed}
\usepackage{tcolorbox}
\usepackage{lipsum}
\usepackage{url}
\usepackage{xcolor}
\usepackage{float}
\usepackage{multicol}
\usepackage[T1]{fontenc}
\usepackage[switch]{lineno}
\usepackage[utf8]{inputenc}
\usepackage[english]{babel}
\usepackage{hyperref}
\hypersetup{
    colorlinks=true,
    linkcolor=blue,
    anchorcolor=blue,
    filecolor=blue,      
    citecolor=blue,
    urlcolor=blue
}

\def\BibTeX{{\rm B\kern-.05em{\sc i\kern-.025em b}\kern-.08em
    T\kern-.1667em\lower.7ex\hbox{E}\kern-.125emX}}

\graphicspath{ {./figure/} }

\begin{document}

\title{A Preliminary Study on Self-Contained Libraries \\ in the NPM Ecosystem\\
}

\author{
\IEEEauthorblockN{Pongchai Jaisri, Brittany Reid, Raula Gaikovina Kula}
\IEEEauthorblockA{\textit{}
\textit{Nara Institute of Science and Technology (NAIST)}\\
Nara, Japan \\
Email: \{jaisri.pongchai.js3, raula-k\}@is.naist.jp,  brittany.reid@naist.ac.jp}
}


\maketitle

\begin{abstract}
The widespread of libraries within modern software ecosystems creates complex networks of dependencies. These dependencies are fragile to breakage, outdated, or redundancy, potentially leading to cascading issues in dependent libraries. One mitigation strategy involves reducing dependencies; libraries with zero dependencies become to self-contained. This paper explores the characteristics of self-contained libraries within the NPM ecosystem. Analyzing a dataset of 2763 NPM libraries, we found that 39.49\% are self-contained. Of these self-contained libraries, 40.42\% previously had dependencies that were later removed. This analysis revealed a significant trend of dependency reduction within the NPM ecosystem. The most frequently removed dependency was babel-runtime. Our investigation indicates that the primary reasons for dependency removal are concerns about the performance and the size of the dependency. Our findings illuminate the nature of self-contained libraries and their origins, offering valuable insights to guide software development practices.

\end{abstract}


\section{Introduction}
In software development, a software library refers to pre-built, reusable code modules or libraries that
developers integrate into their projects to enhance functionality processes. The term dependency in the context of software libraries refers to the reliance of the main library on external modules to function correctly. These dependencies are crucial in ensuring the proper execution of the software by providing essential functionality. Nowadays, the widespread of libraries within modern software ecosystems creates complex networks of dependencies (e.g., NPM for JavaScript, PyPI for Python, and Maven for Java).

The practice of adding numerous dependencies by library maintainers can lead to dependency bloat \cite{10006863}. This presents a challenge, as maintainers may lack clear visibility into which specific parts of the library utilize each dependency. Consequently, the risk associated with dependency usage may increase. The dependencies introduce several challenges. First, they exhibit fragility; complex inter-dependencies mean a single broken dependency can have cascading effects \cite{kula2017impact, venturini2023depended}. Second, dependencies are outdated or abandonment \cite{wattanakriengkrai2022giving}. Third, redundancy is prevalent. The lack of strict publishing guidelines within the ecosystem allows developers to create and distribute libraries with overlapping functionality \cite{chen2021helping, abdalkareem2017developers, abdalkareem2020impact}. Finally, dependencies can potentially lead to cascading issues in dependent libraries.

Software ecosystems present a vast landscape of potential research topics. Directly related to the focus of this paper are investigations into trivial packages, software reuse \& dependency changed, and self-contained library.

One mitigation strategy is dependency reduction. Rather than addressing problems rooted in external dependencies, maintainers may opt to remove problematic libraries entirely. Maintainers have the flexibility to remove dependencies as needed. If the process of dependency removal continues until a library has zero dependencies, it becomes a self-contained library.

In this paper, we further classify self-contained libraries into two distinct categories base on the dependency history:
\begin{enumerate}
    \item Always Self-Contained Libraries: The libraries that maintained self-contained status from their initial release to the latest version (e.g., get-stdin).
    \item Become Self-Contained Libraries: The libraries that initially possessed dependencies but subsequently removed them (e.g., prettier).
\end{enumerate}

\begin{figure*}[h]
    \centering
    \includegraphics[width=\textwidth]{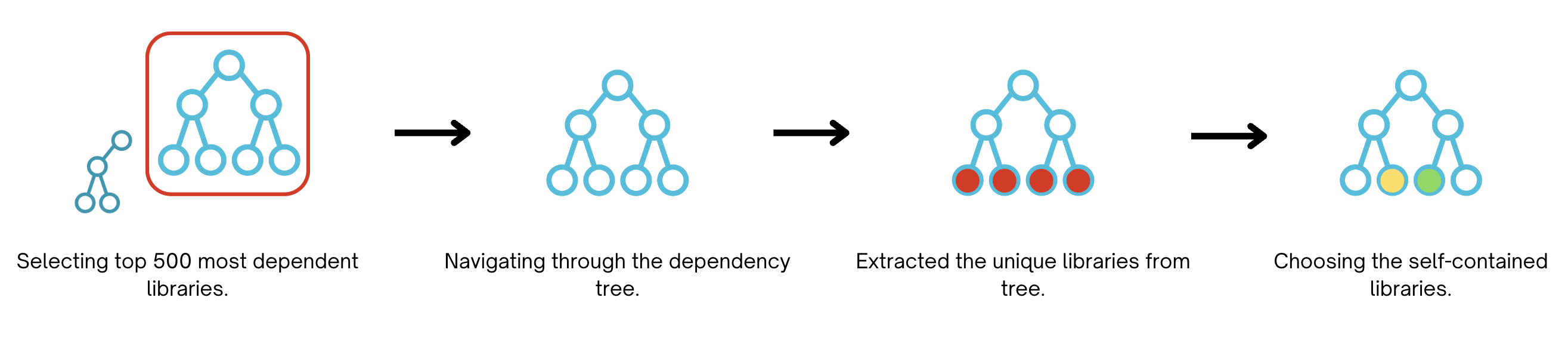}
    \caption{Overview of library selection}
    \label{fig:method}
\end{figure*}

Our idea in this paper is exploring the characteristic of self-contained library within NPM ecosystem, guideline with 3 research questions.
\begin{enumerate}
    \item RQ 1: \textit{To what extent is the reducing of dependencies a prevalent phenomenon?}
        \\ Motivation: Answering this research question will quantify the prevalence of dependency reduction among libraries. This metric will provide insight into the popularity of self-contained libraries and their potential impact within the broader software ecosystem.
    \item RQ 2: \textit{Which types of libraries are most reduced?}
        \\ Motivation: This research question aims to identify the specific types of dependencies that are most frequently removed. Understanding these trends will shed light on evolving practices in dependency management and illuminate potential implications for dependent libraries.
    \item RQ 3: \textit{What factors are frequently associated with libraries that have reduced?}
        \\ Motivation: This question extends the findings of RQ 1 and RQ 2., establishes the prevalence of self-contained libraries, while RQ 2 reveals the dependencies most frequently removed. By synthesizing this data, we can begin to identify common characteristics among libraries that shed dependencies, ultimately leading to a better understanding of the factors influencing this phenomenon.
\end{enumerate}

This study investigates the characteristics of self-contained libraries within the NPM ecosystem. We analyze a dataset of 2,763 NPM libraries to identify the prevalence of libraries that have undergone dependency reduction, transforming them into self-contained libraries. Notably, our findings reveal that 40.42\% of self-contained libraries had dependencies that were subsequently removed. We further explore the most frequently removed dependencies (e.g., \texttt{babel-runtime}) and delve into the motivations behind dependency removal by examining associated Git commits. By analyzing these reasons, we identify the most prevalent factors driving dependency reduction. The most prevalent factors are concerns about the performance and the size of the dependency. This research enhances our understanding of the evolving relationship between dependency libraries and their dependents within the NPM ecosystem, shedding light on dependency management trends.

\begin{figure*}[ht]
    \centering
    \includegraphics[width=\textwidth]{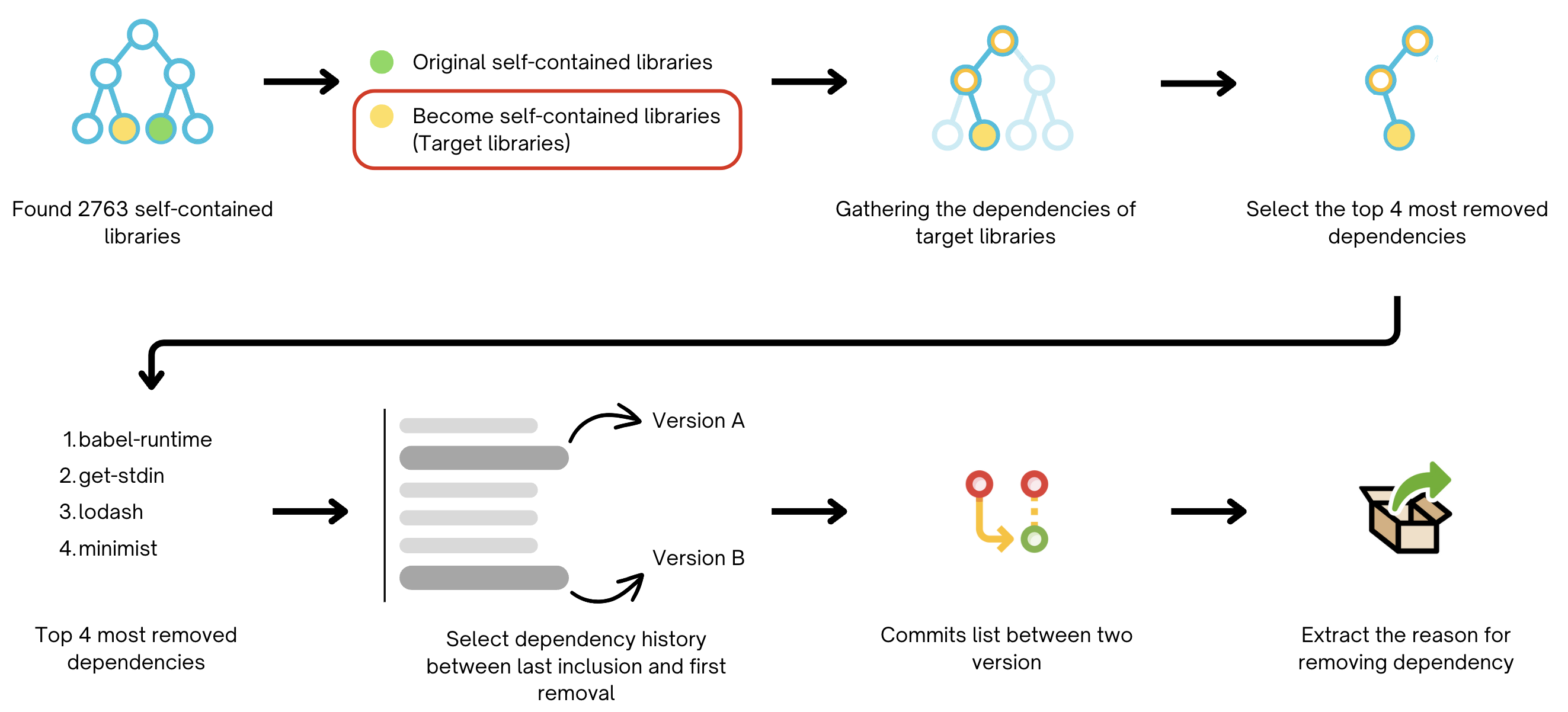}
    \caption{Overview of gathering target commits}
    \label{fig:method_2}
\end{figure*}


\section{Related work}
\subsection{Trivial packages}
Within the NPM ecosystem, the use of small, often single-function, libraries has been observed~\cite{kula2017impact, abdalkareem2017developers}. These so-called `trivial packages' come with several disadvantages. Maintaining a large number of small, trivial packages can significantly increase a developer's workload. Additionally, these packages can lead to complex dependency chains that are difficult to manage and resolve (often referred to as "dependency hell") \cite{abdalkareem2017developers, abdalkareem2020impact}. The abundance of trivial packages can make it challenging to find the right one, especially when multiple packages offer similar functionalities. This overabundance also contributes to redundant packages within the ecosystem. Finally, projects that incorporate many trivial packages tend to have longer installation and build times due to the increased number of dependencies.

\subsection{Software reuse \& dependency changes}
Software reuse is the process of creating software system from existing software rather than building them from scratch \cite{sametinger1997software}. Software reuse is a fundamental practice in software development, offering benefits such as improved quality and reduced effort. This practice involves a provider who creates reusable software components and users who integrate these components, or dependencies, into their own software.

Dependency changes are a common occurrence in software development and can sometimes introduce problems for dependent libraries. Breakage in dependent libraries can occur due to various reasons \cite{venturini2023depended}, including: feature modifications, incompatible provider versions, changes in object types, undefined objects, incorrect code semantics, failed provider updates, function renaming, or missing files.

Dependency change issues extend beyond the NPM ecosystem. For example, a study examining deprecation in Javadoc \cite{sawant2018features} involved the manual analysis of 374 deprecated methods across four major Java APIs to determine whether deprecation reasons were documented.

This work examines how self-contained libraries form. Previous research has focused on removing dependencies~\cite{chuang2022removing}. However, increased dependency removal creates greater potential for libraries to become self-contained.

\section{Dataset Preparation}
Figure \ref{fig:method} provides a visual overview of the library selection process for the dataset. We begin by selecting the top 500 libraries with the highest number of dependents from libraries.io\footnote{\url{https://libraries.io/}}, sorting them in descending order by dependent count. We then systematically analyze the dependency tree of each library by using the Open Source Repository and Dependency Metadata from libraries.io, yielding the dependency trees of 500 libraries. The dataset is available in Zenodo, at \url{https://doi.org/10.5281/zenodo.10972337} \cite{dataset}

\section{Empirical Study}
In this section, we outline our approach to analyzing self-contained libraries, with a focus on understanding the motivations for removing dependencies. Our study is divided into three parts, each strategically designed to address a specific research question:

\subsection{RQ 1: To what extent is the reducing of dependencies a prevalent phenomenon?}
We investigated the dependency histories of self-contained libraries, utilizing the registry npmjs API\footnote{\url{https://registry.npmjs.org}} to collect historical dependency data. We focus exclusively on regular dependencies, excluding both devDependencies and peerDependencies. The devDependencies are necessary only during development and not included in production environments. The peerDependencies, while indicating potential compatibility with other libraries, may not have direct API interactions. To maintain clarity in this investigation, we limit our analysis to regular dependencies, avoiding the complexities introduced by other dependency types.

\subsection{RQ 2: Which types of libraries are most reduced?}
We conducted a detailed analysis of what dependencies were removed from the identified libraries. Utilizing the dependency history data provided by the registry.npmjs API, we compared dependency lists across versions. Our focus was specifically on versions where dependencies were reduced.

\subsection{RQ 3: What are the reasons for reducing libraries?}
Figure \ref{fig:method_2} provides a visual overview of gathering the Git commits of the libraries. For each dependent library, we examined its GitHub\footnote{\url{https://github.com/}} repository to pinpoint the version where the maintainer reduced the target dependency. We collected relevant commits within that period. By matching pull requests with corresponding commits, we analyzed discussions within the pull requests and the associated code changes. Our goal was to identify the primary motivations driving the maintainer's decision to remove the target dependency.

\begin{table}[t]
    \centering
    \caption{Statistic of the libraries and dependencies.}
    \begin{tabularx}{0.75\linewidth} {l|r}
     \toprule
     \textbf{Library categories} & \textbf{No. libraries (\%)} \\ 
     \midrule
     Non Self-contained libraries & 1,672 \hspace{0.75em} (60.51\%) \\
     Self-contained libraries & 1,091 \hspace{0.75em} (39.49\%) \\ 
     \hspace{0.75em} - Become self-contained & 441 \hspace{1.25em} (15.96\%) \\
     \hspace{0.75em} - Original self-contained & 650 \hspace{1.25em} (23.53\%) \\
     \midrule
     All & 2,763 \hspace{1.5em} (100\%) \\
     \bottomrule
    \end{tabularx}
    \label{tab:Statistic}
\end{table}


\section{Result}
\subsection{RQ 1: Prevalence}
    Our analysis of the 500 most depended upon libraries revealed 2,763 unique libraries within the dependency chain. Further investigation into these libraries identified 1,091 as having no dependencies, or self-contained. Of the self-contained libraries, 441 achieved self-contained status through dependency removal. Table \ref{tab:Statistic} provides a detailed statistical breakdown of these libraries and their dependencies.
    
    \begin{tcolorbox}[title=Finding of RQ 1]
        \textbf{We found 39.49\% of unique libraries are self-contained libraries. Moreover, we found 15.96\% of unique libraries became self-contained libraries; that is, they reduced dependencies to zero.}
    \end{tcolorbox}
        
\subsection{RQ 2: Most Reduced Libraries}
\begin{figure}[h]
        \centering
        \includegraphics[width=0.9\linewidth]{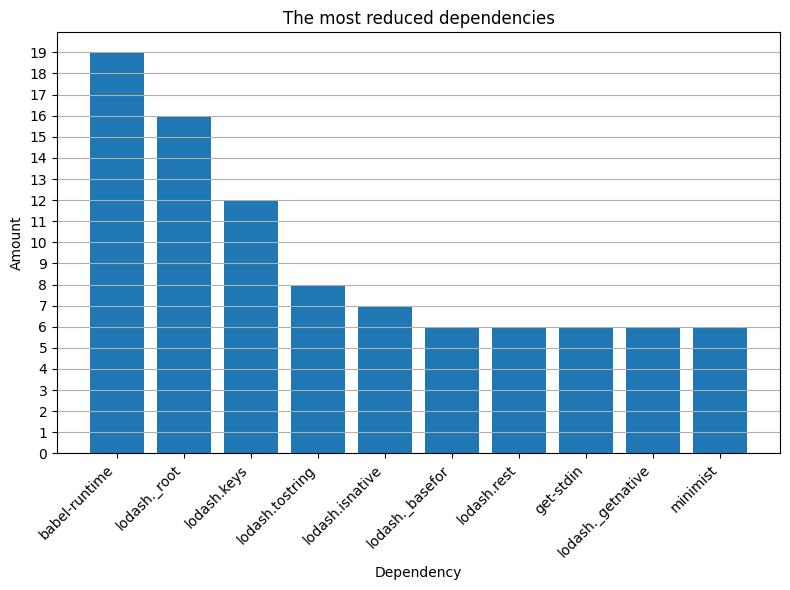}
        \caption{The most reduced dependencies}
        \label{fig:result}
    \end{figure}

    From the 441 libraries that become self-contained libraries, we identified 407 dependencies that were removed. The most frequently removed dependencies were \texttt{babel-runtime}, \texttt{lodash.\_root}, and \texttt{lodash.keys}. Figure \ref{fig:result} reveals a significant trend in dependency reduction, highlighting the most commonly removed dependencies.
    
\begin{figure}[t]
        \centering
        \includegraphics[width=0.9\linewidth]{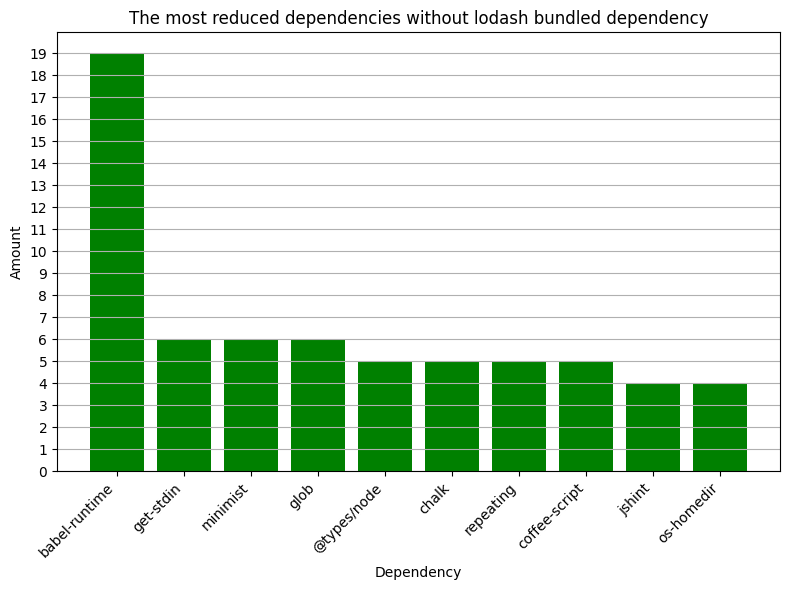}
        \caption{The most reduced dependencies without lodash bundled dependency}
        \label{fig:result_without}
    \end{figure}

\begin{figure}[h]
        \centering
        \includegraphics[width=0.9\linewidth]{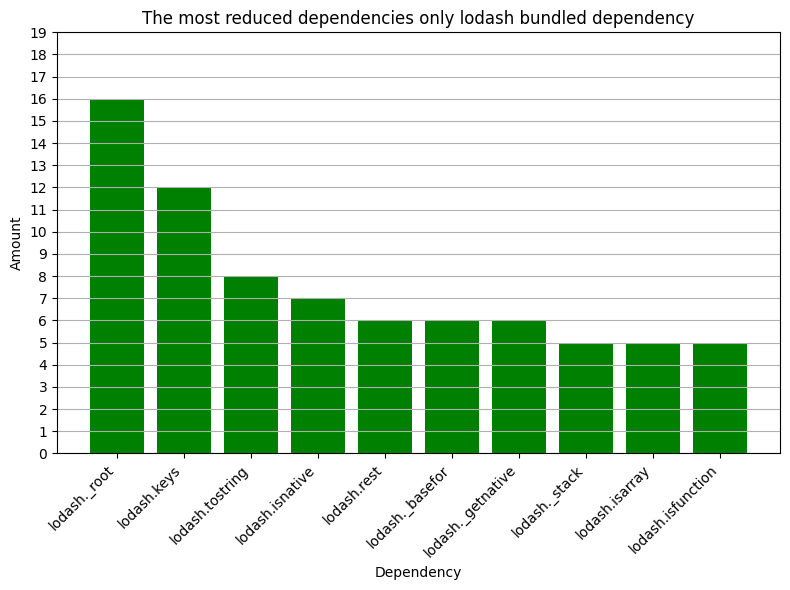}
        \caption{The most reduced dependencies only lodash bundled dependency}
        \label{fig:result_with}
    \end{figure}

    Further analysis revealed that many of these reduced dependencies were, in fact, sub-dependencies (such as \texttt{lodash.\_root} and \texttt{lodash.keys}) from the bundled library \texttt{lodash}. We identified 87 sub-dependencies within the complete list of 407 reduced dependencies. After isolating these sub-dependencies, our revised analysis indicates that the top three most frequently removed dependencies are \texttt{babel-runtime}, \texttt{get-stdin}, and \texttt{minimist}. Figure \ref{fig:result_without} displays the most frequently removed package-type dependencies (excluding sub-dependencies) and the corresponding number of reductions and Figure \ref{fig:result_with} specifically details the most frequently removed \texttt{lodash} sub-dependencies.

    \begin{tcolorbox}[title=Finding of RQ 2]
        \textbf{The most reduced package dependency is \texttt{babel-runtime} while the most reduced sub-dependency is \texttt{lodash.\_root}.}
    \end{tcolorbox}


\begin{table}[h]
\centering
\caption{Number of the gathered commits.}
\begin{tabularx}{0.9\linewidth} {l r r}
    \toprule
    \textbf{Removed dependency} & \textbf{No. commits} & \textbf{No. dependent} \\
    & & \textbf{libraries} \\
    \midrule
    Single dependency & 300 & 27 \\ 
    \hspace{0.75em} - babel-runtime     & 100 & 6 \\
    \hspace{0.75em} - get-stdin         & 100 & 15 \\
    \hspace{0.75em} - minimist          & 100 & 6 \\
    \midrule
    Bundled dependency        & 100 & 1 \\
    \hspace{0.75em} - lodash            & 100 & 1 (7 sub-dependency) \\
    \midrule
    All & 400 & 28 \\
    \bottomrule
\end{tabularx}
\label{tab:Statistic rq 3}
\end{table}

\begin{table*}[h]
    \centering
    \caption{Categories of classifying reasons for 28 dependent libraries.}
    \begin{tabularx}{0.95\linewidth} {l|l|l}
     \toprule
     \textbf{Terminology} & \textbf{Definition} & \textbf{Libraries with commits} \\ 
     & & \textbf{for each terminology (\%)} \\
     \midrule
     Performance of the dependency & The dependency is too heavy, the maintainer of dependents library need to & 5 (17.86\%) \\
     & remove this dependency. \\
     Function replacement & Replace the dependency's function with built-ins function or custom function & 3 (10.71\%) \\ 
     Dependency replacement & Replace the dependency with another dependency & 1 (3.57\%) \\
     Minimize the dependency & The dependency is too heavy, the maintainer of dependency prefer to split & 4 (14.29\%) \\
     & some features of the dependency to independent library. \\
     Removing unused dependency & Remove unnecessary or unused dependency & 1 (3.57\%) \\
     Others & The other reasons that cannot classify & 14 (50\%) \\
     \bottomrule
    \end{tabularx}
    \label{tab:terminology}
\end{table*}
        
\subsection{RQ 3: Reasons for Reducing}
    Guided by the findings of RQ 2, we focus exclusively on reduced dependencies that are not sub-dependencies. Our analysis centers on the top three most frequently removed dependencies: \texttt{babel-runtime}, \texttt{lodash.\_root}, and \texttt{lodash.keys}, looking at 27 parent libraries. Consistent with our methodology, we distinguish between sub-dependencies from bundle dependencies and single dependenc. As in the Table \ref{tab:Statistic rq 3}, we examine 300 random commits from 27 dependent libraries where these three dependencies were reduced (normal dependencies), along with 100 random commits from 7 sub-dependency where bundled dependency were removed. Therefore, these are the top three most frequently removed dependencies after we split sub-dependencies from bundled library. We split 100 commits for each dependency. Below are the reason for the removal of dependencies that we found from random Git commit messages from dependent libraries.
    
    Table \ref{tab:terminology} presents the terminology, definitions used to classify reasons for reducing software dependencies and the overall library with commit for each terminology. This classification criteria was developed through an analysis of commit messages, discussions within matched pull requests, and the relevant code modifications.

    Results for the 28 libraries show that dependency removal was motivated by several factors: performance of the dependency, affecting 5 libraries (18.52\%), function replacement affecting 3 libraries (11.11\%), dependency replacement affecting 1 library (3.7\%), minimalist the dependency affecting 4 libraries (14.81\%), and removing unused dependencies affecting 1 library (3.7\%), the others are 14 libraries (50\%).
    
\noindent\textit{babel-runtime}

We found the reason for dependency-related changes from 3 distinct parent packages: \texttt{babylon}, \texttt{common-tags}, and \texttt{babel}.
\begin{enumerate}
    \item babylon\footnote{\url{https://github.com/babel/babylon/pull/110}}: We found that the maintainer removed \texttt{babel-runtime} because they were using loose mode: a  special mode from babel which helps to refactor the source code. \textit{"remove babel-runtime dep/transform-runtime since we are using loose mode"}. We conclude that this reason is a replacement function.
    
    \item common-tags\footnote{\url{https://github.com/zspecza/common-tags/pull/148}}: we found a few reasons that the maintainer removed \texttt{babel-runtime} because this dependency is a heavy dependency. \textit{"The issue in the article is with @angular/cli using 3 methods from common-tags, but the outlaying issue is babel-runtime is a heavy dependency."}. For this reason, we conclude that this reason is a performance of the dependency. The another reason is the maintainer replace the babel-runtime with Typescript which can handle the browser compatibility. \textit{"We are considering dropping Babel and replacing it with Typescript, which would inline all the needed stuff. We care about browser compatibility, which is what Babel is used for, and Typescript is handling that well (even as far as ES3)."}. We conclude that this reason is a dependency replacement.
    
    \item babel\footnote{\url{https://github.com/babel/babel/issues/5118}}: We found variety of reasons that the maintainer removed \texttt{babel-runtime}.
    \begin{itemize}
        \item The first reason is they want to replace function from \texttt{babel-runtime} with built-ins function from the new version of Node. \textit{"The reason for doing this in the first place has to do with wanting to use built-ins like Symbol/Promise, etc which are not native to node 0.10/0.12. Now that we are on >= Node 4 we should be able to use the native ones."}. We conclude that this reason is a function replacement.

        \item The second reason is babel is a big dependency because deduping never works and the maintainer encountered with many copies of \texttt{babel-runtime}. \textit{"Babel is always the biggest dependency in my projects because deduping never works and I end up with a gadzillion copies of babel-runtime."}. We conclude that this reason is a performance of the dependency.

        \item The third reason is the maintainer thinking about less dependency and use native will make \texttt{babel} faster to run and uninstall.\textit{"less dependencies, use native, might be faster to run/install"}. We conclude that this reason is a performance of the dependency.
    \end{itemize}
    
\end{enumerate}
    
\noindent\textit{get-stdin}

We found the reason from 5 difference dependent packages; \texttt{dateformat}, \texttt{pretty-bytes}, \texttt{indent-string}, \texttt{detect-indent}, and \texttt{detect-newline}.
\begin{enumerate}
    \item dateformat\footnote{\url{https://github.com/felixge/node-dateformat/issues/36}}: We found that the maintainer removed \texttt{get-stdin} with other libraries. They concern about the flattened dependency tree of the dateformat. \textit{"I love using this, but I just realized that it's using meow, and I'm guessing that a good percentage of users like myself probably aren't using the CLI, but the dependency tree for it is absurdly massive. Especially considering this would have zero deps without it. Meow has 48 dependencies in total, this is its flattened dependency tree."}. We conclude that this reason is a performance of the dependency.
    
    \item pretty-bytes \cite{pretty_bytes}, indent-string \cite{indent-string}, detect-indent \cite{detect_indent}, and detect-newline \cite{detect_newline}: We found that these libraries have a same maintainer. The maintainer of these module split the function in the libraries into normal version and CLI version. The normal version used to have dependencies but the maintainer separate them into the CLI version. We conclude that this reason is a minimalist the dependency.
\end{enumerate}

\noindent\textit{minimist}

We found the reason for dependency removal from 6 difference dependent packages; \texttt{flow-parser}, \texttt{envinfo}, \texttt{prettier}, \texttt{indent-string}, \texttt{detect-indent}, and \texttt{detect-newline}.
\begin{enumerate}
    \item flow-parser\footnote{\url{https://github.com/facebook/flow/pull/3586/files}}: The maintainer remove function name "flowparse" and "flowvalidate" from the library. Therefore, they removed the dependency which support those function. We conclude that the reason is a removing unused dependency.

    \item envinfo\footnote{\url{https://github.com/tabrindle/envinfo/pull/29}}: The maintainer relocated minimist from regular dependencies to devDependencies, meant for development-only packages that are removed during the build process. We conclude that the reason is a performance of dependency.

    \item prettier\footnote{\url{https://github.com/prettier/prettier/pull/1850}}: We found that the maintainer remove \texttt{minimist} because they want to transform \texttt{prettier} to dependency-free. \textit{"Move all the dependencies to dev dependencies and --exact. Since we are now bundling all the dependencies, we can have prettier be dependency-free on npm <3"}. We conclude that this reason is a function replacement.

    \item indent-string, detect-indent, and detect-newline: The reason is the same reason with removing \texttt{get-stdin} which is a minimalist the dependency.
\end{enumerate}

\noindent\textit{lodash}

Among removed dependencies classified as sub-dependencies from bundle dependencies, the top three most frequent were \texttt{lodash.\_root}, \texttt{lodash.keys}, and \texttt{lodash.tostring}. However, the maintainer did not provide explicit reasons for the removal of these specific sub-dependencies.



\begin{tcolorbox}[title=Finding of RQ 3]
    \textbf{The primary factor for removing dependency is \textit{'the performance of the dependency'}, affecting more than 18\% of the dependent libraries. This is followed by \textit{minimize the dependency} (more than 14\%) and the \textit{function replacement} (more than 11\%)}.
\end{tcolorbox}

\section{Discussion and Future work}
In this work, we show that developers may be changing their attitudes on simply adopting blindly adopting dependencies into their applications. Based on the results of the study, one potential application is the development of a tool that recommends dependencies suitable for removal. Informed by the characteristics of frequently removed dependencies identified in this work, such a tool could analyze dependency lists (e.g., package.json in JavaScript, pyproject.toml in Python, etc.) and provide insights regarding individual dependencies. A tool of this nature could prove valuable to a broad range of users within the development community, extending beyond the immediate maintainers of self-contained libraries.

\section*{Acknowledgement}
This work is supported by the Japanese Society for the Promotion of Science (JSPS) KAKENHI Grant Number JP20H05706.

\bibliographystyle{ieeetr}
\bibliography{ref.bib}

\end{document}